\documentclass[conference]{IEEEtran}
\IEEEoverridecommandlockouts
\usepackage{cite}
\usepackage{amsmath,amssymb,amsfonts}
\usepackage{algorithmic}
\usepackage{graphicx}
\usepackage{textcomp}
\usepackage{xcolor, soul}
\sethlcolor{yellow}
\usepackage{hyperref}
\usepackage{tikz}
\usepackage{stfloats}
\usepackage{placeins}
\def\BibTeX{{\rm B\kern-.05em{\sc i\kern-.025em b}\kern-.08em
    T\kern-.1667em\lower.7ex\hbox{E}\kern-.125emX}}
    
\makeatletter
\newcommand{\linebreakand}{%
  \end{@IEEEauthorhalign}
  \hfill\mbox{}\par
  \mbox{}\hfill\begin{@IEEEauthorhalign}
}
\makeatother

\begin{document}

\title{Predicting Stock Market time-series data using CNN-LSTM Neural Network model\\
}

\author{\IEEEauthorblockN{Aadhitya A}
\IEEEauthorblockA{\textit{Department of Information Technology} \\
\textit{Madras Institute of Technology}\\
Chennai, India \\
Email: aadhitya864@gmail.com}
\and
\IEEEauthorblockN{Rajapriya R}
\IEEEauthorblockA{\textit{Department of Information Technology} \\
\textit{Madras Institute of Technology}\\
Chennai, India \\
}
\and
\IEEEauthorblockN{Vineetha R S}
\IEEEauthorblockA{\textit{Department of Information Technology} \\
\textit{Madras Institute of Technology}\\
Chennai, India \\
}
\linebreakand
\IEEEauthorblockN{Anurag M Bagde}
\IEEEauthorblockA{\textit{Department of Information Technology} \\
\textit{Madras Institute of Technology}\\
Chennai, India \\
}
}

\maketitle

\begin{abstract}
Stock market is often important as it represents the ownership claims on businesses. Without sufficient stocks, a company cannot perform well in finance. Predicting a stock market performance of a company is nearly hard because every time the prices of a company’s stock keeps changing and not constant. So, it’s complex to determine the stock data. But if the previous performance of a company in stock market is known, then we can track the data and provide predictions to stockholders in order to wisely take decisions on handling the stocks to a company. To handle this, many machine learning models have been invented but they didn't succeed due to many reasons like absence of advanced libraries, inaccuracy of model when made to train with real time data and much more. So, to track the patterns and the features of data, a CNN-LSTM Neural Network can be made. Recently, CNN is now used in Natural Language Processing (NLP) based applications, so by identifying the features from stock data and converting them into tensors, we can obtain the features and then send it to LSTM neural network to find the patterns and thereby predicting the stock market for given period of time. The accuracy of the CNN-LSTM NN model is found to be high even when allowed to train on real-time stock market data. This paper describes about the features of the custom CNN-LSTM model, experiments we made with the model (like training with stock market datasets, performance comparison with other models) and the end product we obtained at final stage.
\end{abstract}


\section{Introduction}
Stock Market (or share market) is a place where corporate folks or businessmen often look on their "shares" or ownership claims related to the company/organisation. Stock market is considered an important place in economy as a country's wealth is decided on it. The hardest task in stock market is predicting it because the data is not constant, always. In early days, many people tried to predict stocks using conventional methods but most of the time, failed. So the probability of predicting stocks correctly is very minimal. Today, Machine Learning (ML) techniques have been implemented to predict the company shares and thereby providing suggestions to stock holders to improvise the company's financial growth. But they are not accurate. This paper focuses on some of the works done in predicting stock market and a new method to follow CNN-LSTM Neural Network model approach to predict data for given time series data.

\section{Related Work}

Before the time of writing this paper, many have proposed and implemented various algorithms in order to predict stock market data. We did a literature survey to find some of the algorithms proposed and found some of the advantages, disadvantages present in those algorithms. Subhadra and Kalyana in \cite{b1} conducted analysis in various machine learning methods to predict stock market data and found that Random Forest performs good compared with Linear Regression and other algorithms, however the error percentage rises in the model when the input data is not smoothed in pre-processing stage. Pang, Zhou et al in \cite{b2} made comparisons with RNN and LSTM model and conducted experimental analysis, in which LSTM with Auto-Encoder module enabled (AELSTM) predicted most of the data but the implementation is done using old libraries and with real-time stock market data, thus the accuracy of the model was low. This was improved but revealed an important point when a model is made to train with real-time data. Uma and Kotrappa in \cite{b3} proposed a new method where LSTM with Log Bilinear layer on top of it. The model predicted most of the stock market data and turned out with high accuracy but it was proposed and not tested with real time data, also it was meant to predict data only during the time of COVID-19 and not beyond that. Kimoto, Yoda, Takeoka in \cite{b4} discussed a buying and selling timing prediction system based on modular neural network which converts the technical indexes and economic indexes into a space pattern to input to the neural networks. During the analysis phase, neural network model produced a higher correlation coefficient in comparison to multiple regression. The experiment did by Guresen, Kayakutlu and Daim in \cite{b5} evaluates the efficiency of dynamic artificial neural network (DAN2), multi-layer perceptron (MLP), and hybrid neural networks. The paper concludes by stating that the classic ANN model - MLP gives the most reliable results in forecasting time series while Hybrid methods failed to improve the forecast results. Nelson, Pereira and Oliveria in \cite{b6} proposed an LSTM network to predict future trends of stock prices in time steps of 15 minutes based on the price history, alongside technical analysis indicators. On average  55.9\% accuracy was achieved in predicting whether the price of a particular stock may increase or not shortly in the future. Selvin, Menon, Soman et al in \cite{b7} experimented on three different deep learning models, namely CNN, RNN, and LSTM with a sliding window approach. Out of the three, CNN gave more accurate results than the other two models which is due to the reason that CNN uses only the information on the current window for predicting stock price. This allows CNN  to understand the dynamic changes and patterns occurring in the current window. Conversely, RNN and LSTM use information from previous lags for predicting future instances. Hiransha, Gopalakrishnan and Soman in \cite{b8} conducted experiments to compare different Deep Learning models viz. ANN, MLP, LSTM, and RNN. ANN captured the pattern at the initial stage so did RNN but on reaching a certain time period both failed to identify the pattern. Same was the case with LSTM but CNN seemed to perform better compared to the other three networks even though several time periods showed less accuracy for the predicted values. Bansal, Hasija et al in \cite{b9} proposed an Intelligent decentralized Stock market model using the convergence of machine learning alongside DAG-based cryptocurrency. The ML model which is based on LSTM achieved an accuracy of 99.71\%  in prediction. The feature vector of stock for the company contained 4 parameter values i.e. ‘open’, ‘close’, ‘low’, and ‘high’ with batch size as 50 for 100 epochs.  

\begin{table*}[!htbp]
	\caption{Summary of work did by researchers}
	\begin{center}
		\resizebox{\textwidth}{!}{
		\begin{tabular}{|c|c|c|c|}
			\hline
			\multicolumn{4}{|c|}{\textbf{Summary of research works}} \\
			\cline{1-4} 
			\textbf{\textit{Reference}}& \textbf{\textit{Method}}& \textbf{\textit{Stock Market/Dataset used}} & \textbf{\textit{Metrics used}} \\
			\hline
			\cite{b1} & Random Forest, Linear Regression & Customized dataset & Variance, MSE, MAE \\
			\hline
			\cite{b2} & LSTM, ELSTM, AELSTM & Shanghai A-Share composite index, Sinopec & MSE, Accuracy  \\
			\hline
			\cite{b3} & LBL-LSTM & Mixed & MSE, Accuracy \\
			\hline
			\cite{b4} & Neural Network & TOPIX (Japan) & Correlation Coefficient \\
			\hline
			\cite{b5} & ANN, DAN2, MLP, Hybrid NN & NASDAQ & MSE, MAD, Coefficient score \\
			\hline
			\cite{b6} & LSTM & BOVA11, BBDC4, ITUB4, CIEL3, PETR4 & Accuracy, F1, Precision, Recall \\
			\hline
			\cite{b7} & CNN, RNN, LSTM with sliding-window approach & NSE & Error percentage \\
			\hline
			\cite{b8} & ANN, MLP, LSTM, RNN & NSE & MAPE \\
			\hline
			\cite{b9} & LSTM with smart contracts (DAG-based) & NYSE & MSE\\
			\hline
		\end{tabular}
	}
		\label{research1}
	\end{center}
\end{table*}


\section{Proposed Work}
After we conducted the literature survey, we realised some of the key points to be taken while designing a ML model
\begin{itemize}
\item The model must be designed in such a way that it should parse real-time stock market data and not just sample data
\item The data has to be preprocessed correctly in order to avoid errors during training and testing phase
\item The accuracy of the model not only depends upon it's parameters but also the dataset as well
\item The only point most of the papers didn't mention is the deployment, as it's one of the most important module while making an ML/DL model
\end{itemize}
Considering the above key points, we focused on making the ML model. We decided to go on with CNN-LSTM Neural Network (Convolutional Neural Network and Long Term Short Memory Neural Network) approach because CNN helps in tracking the features of dataset and LSTM helps in tracking the patterns, allowing to train on them. This approach isn't the first time as some researchers already tried to implement the CNN-LSTM method but we tweaked the parameters, kernel sizes (for CNN) and layers to experiment and test it on real-time data. Since this is a regression type of problem where we had to train with time-series data, we used Mean Square Error as the standard metric rather than accuracy.

\begin{figure*}[htbp]
\centerline{\includegraphics[scale=0.3]{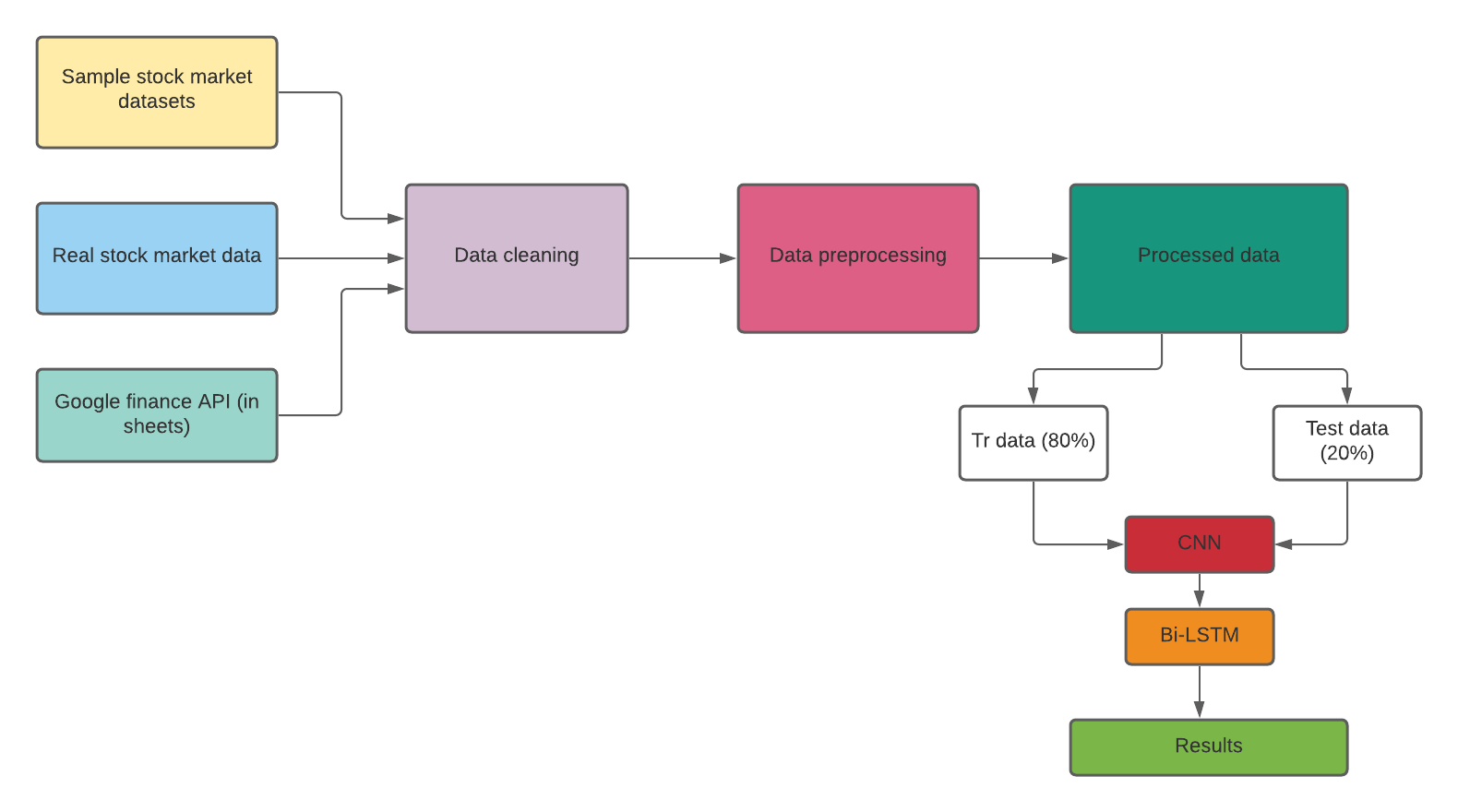}}
\caption{Architecture for Deep Learning model}
\label{model-sam}
\end{figure*}

For the neural network, we first analysed with other works and then decided the architecture in order to maintain novelty. The architecture diagram for the neural network is shown in Fig. \ref{model-sam}.

\textbf{NOTE: For this project, we mainly used Python and Jupyter Notebooks via Colab and Kaggle to perform the experiment. To see the experiment we did, refer \url{https://github.com/Circle-1/Stock-X}}

\section{Experimental Analysis}
For the experiment, we used Python for data preprocessing and creating a Neural Network model. For deployment, it's considered as minimal because, we saved the model in HDF5 format using Keras, but in addition to that, we made a Docker Image and Helm charts for people to work on. In this section, the major parts of the project and the experiments performed are described.

\subsection{Data Collection, Analysis and Preprocessing}\label{A}
Before making a model, the first step is to collect enough datasets such that the base analysis is made to study about the stock market data. So, we gathered enough datasets from Kaggle (explained in later stages) but realised that they are sample ones and we had to search for real time ones. Then, we came across several finance APIs like Yahoo finance, Alpha Vantage which helps in gathering stock data for specific time period. So, we took Alpha Vantage API and used "TIME\_SERIES\_DAILY" option to obtain stock data of a company ranging since 10 years. We used "full" mode to collect enough data rather than using "compact" mode in API (which fetches only 100 columns meant for rapid usage cases) and we were able to collect the data for any company with valid API keys. Some stock data is also gathered using Google Sheets via "GOOGLEFINANCE" function. Then we stored the data in CSV format for testing phase.

Then, we did an Exploratory Data Analysis (EDA) on the dataset to know about the stock market data in depth. We also implemented Moving Average and Daily Return columns to know how a stock market works and analysed some of the features present in it. After that, we went to preprocessing phase.

\begin{figure}[htbp]
\centerline{\includegraphics[scale=0.7]{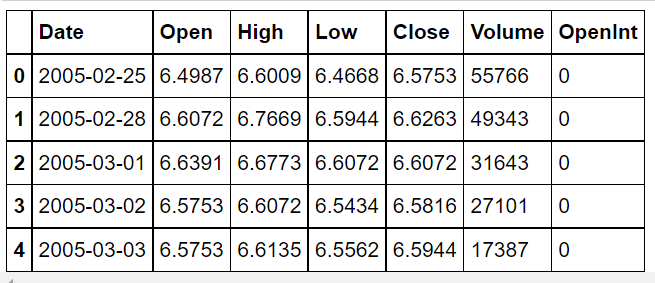}}
\caption{Example of stock market dataset}
\label{dataset-1}
\end{figure}

In preprocessing phase, we first cleansed the data by removing NULL values from the dataset and taking the mean of data and replacing it if necessary using Pandas library. Then we took the four columns of any stock market dataset, namely "Open", "Close", "High", "Low". These are the columns which mainly involve in training the dataset especially "Close" column (shown in Fig. \ref{dataset-1}). The graphs are plotted using the matplotlib and seaborn library in Python. 

During preprocessing stage, we realised that CNN always considers 2-Dimensional and 3-Dimensional arrays to train and select required features. But here, the data we have is of 1-Dimensional arrays. This is one of the reasons why CNN is often seen in Computer Vision (CV) based applications and not in NLP based applications. 

So, in order for CNN model to parse the dataset, we made a function where the 1-D arrays are made to convert to [100,1] tensors (precisely, a vector). Tensors are a type of data structures that describe a mulilinear relationship between set of objects in a vector space. So, for converting 1-D array to tensor, every 100 rows are taken and from that the mean of the values are calculated and made to store in a separate column. This process is done for entire dataset. In our case, we did this on the "Close" column as it's the main column where we would decide the prediction of the stock data.

After this step, we would obtain tensors for CNN side of model to train. Then, we split 80\% for training and 20\% for testing. Finally, we reshaped the data and sent it to the training phase. 

\subsection{Training Phase}\label{B}
After the dataset is processed, the NN model has to be made. In our case, it's the CNN-LSTM Neural Network model. 

For our model, we considered to divide the model into two parts, CNN and LSTM.

\subsubsection{CNN}
For the CNN section of model, we followed a custom way instead of ascending kind of way in the size of layers. So, we made 3 layers of neuron size 64,128,64 with kernel size=3 along with MaxPooling layers in between. Finally we added a Flatten layer at end of CNN section to convert the tensors back to 1-D array.

All CNN layers are added with TimeDistributed function in order to train every temporal slice of input, as we're approaching a Time-Series problem in this case. Then, the processed data is sent to LSTM layers.

\subsubsection{LSTM}
For the LSTM section, we made 2 Bi-LSTM layers to detect the features and train them forward \& backward. For each layer, the neuron size is 100. Additionally, dropout layers are added in between with value of 0.5 in drop some features for stability.

Last, we added a dense layer with linear activation function and at the final layer we used "adam" optimizer ("sgd" also worked in this case but we found "adam" optimizer to be accurate after analysis), Mean Squared Error (mse) as the loss function and "mse" and "mae" as metrics.

\begin{figure*}[htbp]
	\centerline{\includegraphics[scale=0.6]{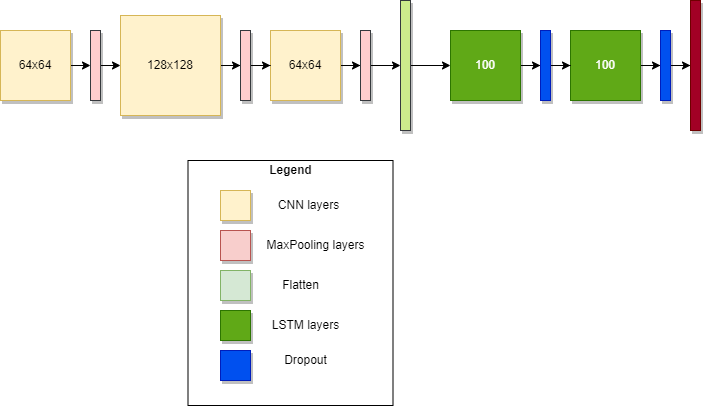}}
	\caption{The architecture of the CNN-LSTM model}
	\label{modelx}
\end{figure*}

\begin{figure}[htbp]
\centerline{\includegraphics[scale=0.5]{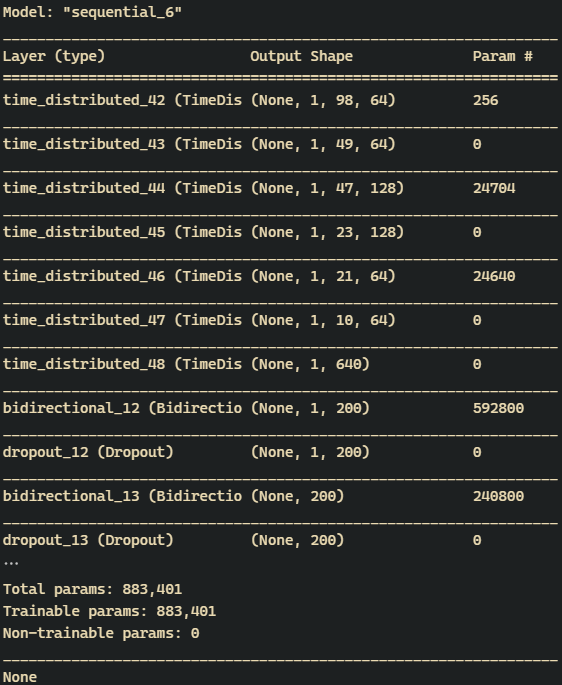}}
\caption{The summary of the proposed model}
\label{model}
\end{figure}

\begin{figure}[htbp]
	\centerline{\includegraphics[scale=0.17]{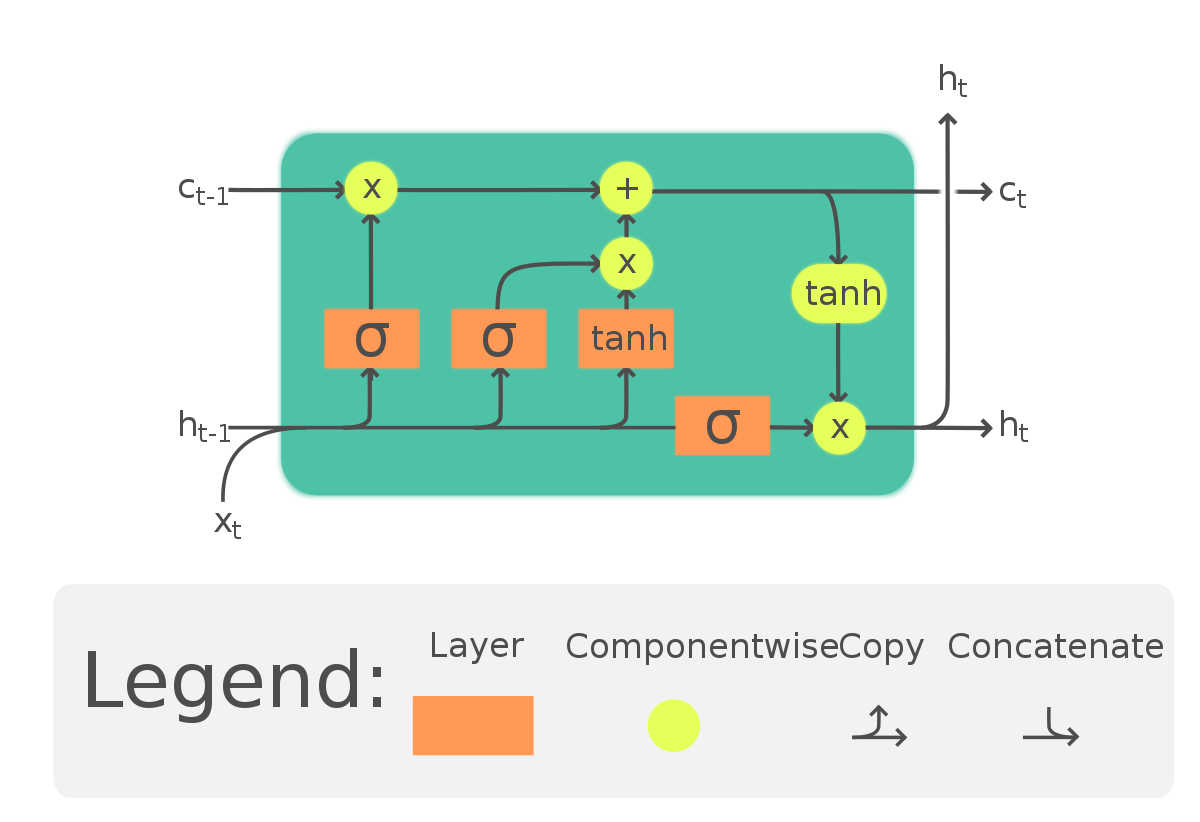}}
	\caption{LSTM Module (Ref: LSTM, Wikipedia)}
	\label{modely}
\end{figure}

\begin{figure}[htbp]
\centerline{\includegraphics[scale=0.7]{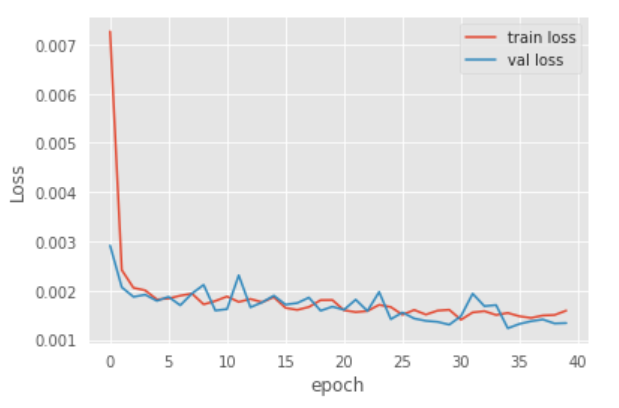}}
\caption{The loss graph obtained during training}
\label{loss-1}
\end{figure}

\begin{figure}[htbp]
\centerline{\includegraphics[scale=0.7]{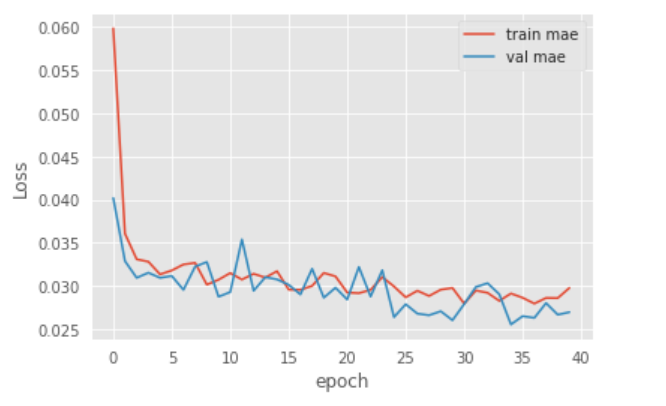}}
\caption{MAE obtained during training}
\label{loss-2}
\end{figure}

\begin{figure}[htbp]
\centerline{\includegraphics[scale=0.7]{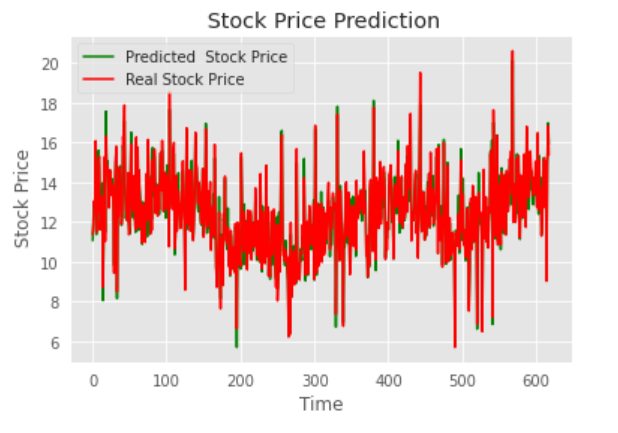}}
\caption{The prediction graph for sample NSE/NASDAQ data (shuffled)}
\label{predict-1}
\end{figure}

The architecture for the model is shown in Fig. \ref{model}. After the model is trained, we made it to plot the graph for the loss values (both training and validation) and at first it proved to be less and later it varied accordingly (shown in Fig. \ref{loss-1} and \ref{loss-2}.). But the loss and mse varies when the model is saved and made to train again with saved parameters (this is fully explained in the testing phase). Then, we managed to refine the test dataset back to arrays using the reshape() function, then made the model to predict the dataset. The graph is plotted and the results proved to be great. The model was able to predict most of the stock data with given time series and it's shown in Fig. \ref{predict-1}. 

\begin{table}[htbp]
	\caption{Accuracy scores}
	\begin{center}
		\begin{tabular}{|c|c|}
			\hline
			\multicolumn{2}{|c|}{\textbf{Scores}} \\
			\cline{1-2} 
			\textbf{\textit{Name}}& \textbf{\textit{Score}} \\
			\hline
			MSE & 0.035 \\
			\hline
			MAE & 0.075 \\
			\hline
			Variance & 0.935370\\
			\hline
			R2 Score & 0.9353 \\
			\hline
			Max Error & 0.174930 \\
			\hline
		\end{tabular}
		\label{tab-mse}
	\end{center}
\end{table}

\subsection{Testing Phase}
After the model has been trained and the values are noted, we saved the model in HDF5 format using Keras API in TensorFlow library. Then, we loaded the HDF5 file and tried training the model again but this time with the different dataset, we were able to train the model but the loss value varies accordingly. For example, if the loss is 0.055 during training phase, the loss increases to 0.153 (estimated, not accurate). It is also found that this happens depending on the dataset we use, for NIFTY sample dataset the error didn't occur whereas in the NASDAQ dataset it occured while loading up the saved model.

\begin{figure*}[htbp]
\centerline{\includegraphics[scale=0.7]{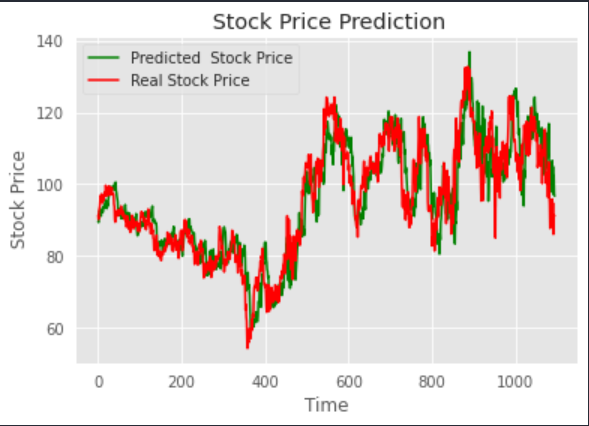}}
\caption{The prediction graph for real stock data [IBM] (un-shuffled)}
\label{predict-2}
\end{figure*}

We tested and experimented the model with different datasets from Kaggle consisting sample data of mixed content from different stock markets \cite{b10}, NIFTY-50 \cite{b11}, NASDAQ and NYSE \cite{b12} to find how the model copes up with different stock market (shown in Fig. \ref{predict-1} and Fig. \ref{predict-2}). We also did with real time dataset by fetching data from Alpha Vantage API (at first step) and tested the model with stock data both shuffled and un-shuffled. It was able to predict most of the stocks as shown in the figures and in table \ref{tab-mse}.

\begin{table*}[htbp]
	\caption{Accuracy with datasets}
	\begin{center}
		\scalebox{1.2}{
		\begin{tabular}{|c|c|}
			\hline
			\multicolumn{2}{|c|}{\textbf{Dataset and MSE Scores}} \\
			\cline{1-2} 
			\textbf{\textit{Dataset}}& \textbf{\textit{MSE Score}} \\
			\hline
			NIFTY (SBIN - sample) & 0.001 \\
			\hline
			NASDAQ (ACTG - sample) & 0.1565 \\
			\hline
			NASDAQ (AAOI - sample) & 0.0016\\
			\hline
			NYSE (IBM - real) & 0.0027 \\
			\hline
			BSE (RELIANCE - real) & 0.0145 \\
			\hline
		\end{tabular}
		}
		\label{tab-mse}
	\end{center}
\end{table*}

\subsection{Deployment}
This section is considered as minimal one because we used Kaggle notebooks in order to create and deploy the models but we realised some developers who don't work with Kaggle and prefer local environments and tools like Conda. So, we made a Docker Image for them in order to work with Jupyterlab (instead of Jupyter notebook) with the GitHub repository as the folder. Initially, it was hard because by default, Docker Images are huge in size because of the OS images like Ubuntu. We got the image size as 250MB at first (without installing any Python libraries) and 750MB when uncompressed. We also tried to use pre-defined images for Jupyter notebooks where the libraries are already installed and ready, but we didn't use them as they are of huge size (approximately 3GB). So we added some libraries and made a docker image with Conda environment (via miniconda) on top of it, still we reached 1GB of data due to the size of libraries and scripts bundled so it's considered to be used only when a user doesn't want to use Kaggle as a working environment.

For deploying the image to Docker Hub, we used GitHub actions which is a CI tool from GitHub to write a CI action in which an image is published every time a change or release is made, so that users need not build image everytime locally. Docker Inc. already made a GitHub action \cite{b13} to perform the above functions, so we used and modified it to build and push Docker images. Similarly, we also made Helm charts if a person needs to run the Docker Image in a Kubernetes cluster. Initially we faced difficulties but after trial and error, we were able to expose the service to the network and run our Docker Image in Kubernetes. The Helm charts are deployed on the Artifact Hub (https://artifacthub.io) and anyone can download the templates from the repository.

\FloatBarrier
\section{End Product}

After several works and tests, we were able to obtain a great working CNN-LSTM Neural Network model which can able to predict most of the stocks when the relevant stock market data is provided. In order to check how our model performed, we compared with other models which are obtained during the literature survey and found the results in the table \ref{tab1}

The model experimented from \cite{b3} is proposed one and haven't been tested rigorously. The model from \cite{b9} performed well and stated that an average of 0.0003 to 0.0004 due to the fact that the data is processed and decentralized. In our case, we obtained an MSE of 0.001 maximum and 0.035 in average as it varied accordingly with different datasets we experimented.

We also experimented our model with different models made from our team* to see how the CNN-LSTM model performed well when compared with other models too. The results were shown in table \ref{tab2}

The varying accuracy of the CNN-LSTM model lies on how the data is processed and the order of the data items (shuffled or un-shuffled). In both cases, the model performed pretty fine and was almost able to predict most of the time-series data with minimal error and variance.

\begin{table}[htbp]
	\caption{Accuracy scores of custom models}
	\begin{center}
		\begin{tabular}{|c|c|}
			\hline
			\multicolumn{2}{|c|}{\textbf{Scores}} \\
			\cline{1-2} 
			\textbf{\textit{Model}}& \textbf{\textit{MSE Score}} \\
			\hline
			CNN-LSTM & 0.035 \\
			\hline
			LSTM & 0.045 \\
			\hline
			XGBoost & 0.047 \\
			\hline
		\end{tabular}
		\label{tab2}
	\end{center}
\end{table}

\begin{table*}[!htbp]
	\caption{Comparison with other models}
	\begin{center}
		\begin{tabular}{|c|c|c|}
			\hline
			\multicolumn{3}{|c|}{\textbf{Models}} \\
			\cline{1-3} 
			\textbf{\textit{Model name}}& \textbf{\textit{Reference}}& \textbf{\textit{Accuracy}} \\
			\hline
			\hl{CNN-LSTM (Our model)} & & MSE = 0.035 (avg)  \\
			\hline
			Random Forest & \cite{b1} & EVS = -0.400594 \\
			\hline
			AELSTM & \cite{b2} & 53\% (Avg) \\
			\hline
			LBL-LSTM (Proposed) & \cite{b3} & 0.017 (Train), 0.026 (Test) \\
			\hline
			LSTM (Decentralized) & \cite{b9} & MSE = 0.0003 \\
			\hline
			IKN-ConvLSTM & \cite{b14} & 98.307\% \\
			\hline
			k-NN Regression & \cite{b15} & 90\% \\
			\hline
			Rider-MBO-based Deep-ConvLSTM & \cite{b16} & MSE = 7.2487, RMSE = 2.6923 \\
			\hline
			SVM \& Logistic Regression & \cite{b17} & 87\% to 90\% \\
			\hline
			SVM with RBF Kernel & \cite{b18} & 89\% \\
			\hline
			LSTM & \cite{b19} & MSE = 3.5 to 3.7 \\
			\hline
			Auditory Algorithm (Proposed) & \cite{b20} & Accuracy is varied (Max: MSE = 0.0002) \\
			\hline
			Fast Fourier Transform model & \cite{b21} & Coefficient = 71.9\% (Average) \\
			\hline
			CPLM-based NARX model & \cite{b22} & RMSE = 17.15 (Average) \\
			\hline
			ANN with Particle Swarm Optimization & \cite{b23} & R-Squared=0.9795 \\
			\hline
		\end{tabular}
		\label{tab1}
	\end{center}
\end{table*}

\FloatBarrier
\section {Conclusion}
Designing a customised CNN-LSTM is bit tricky at first, because there are many algorithms already present to predict the stock data, however not fully optimized. We trained and tested the model with different kinds of datasets like NYSE, NASDAQ and NIFTY and found that accuracy of the model varies accordingly. For NIFTY \cite {b11}, the model proved to be good and predicted at most 99\% of stocks even during the testing stage. But for other datasets like NYSE \cite{b12}, the accuracy varied during the testing phase and not during training. We believe it may be due to the noise present in the dataset during the testing phase and compiling it again. Also, we obtained MSE of value 0.001 to 0.05 (approx) during training and 0.002 to 0.1 (approx) for validation with different datasets. Overall, the paper presents on how CNN can be combined with LSTM to obtain the features from the tensors processed from dataset and detect patterns based on the features obtained. The paper also presents one of the ways in which stock market can be predicted with great accuracy.

\vspace{12pt}

\end{document}